\documentclass[letter,longauth]{aa} % for the letters 
\usepackage{graphicx}
\usepackage{txfonts}
\usepackage[colorlinks=true,linkcolor=blue,citecolor=blue,filecolor=blue,urlcolor=blue]{hyperref}

\begin{document} 
   \title{\vspace{-0.3cm}  Matter power spectrum reconstruction with KiDS-Legacy: Improved internal $\Lambda$CDM consistency and preference for strong baryonic feedback}
   \titlerunning{Matter power spectrum reconstruction with KiDS-Legacy}
   \authorrunning{Broxterman et al.}

   \author{Jeger C. Broxterman
          \inst{1,2}\thanks{broxterman@strw.leidenuniv.nl}
          \and
          Patrick Simon
          \inst{3}\thanks{psimon@astro.uni-bonn.de}
          \and
          Lucas Porth
          \inst{3}
          \and
          Konrad Kuijken
          \inst{2}
          \and
          Angus H. Wright
          \inst{4}
          \and
          Marika Asgari 
          \inst{5}
          \and
          Maciej Bilicki
          \inst{6}
          \and
          Catherine Heymans
          \inst{7,4}
          \and
          Hendrik Hildebrandt 
          \inst{4}          
          \and
          Henk Hoekstra
          \inst{2}
          \and
          Benjamin Joachimi
          \inst{8}
          \and 
          Shun-Sheng Li
          \inst{4,2}
          \and 
          Matteo Maturi
          \inst{9,10}
          \and
          Lauro Moscardini
          \inst{11,12,13}
          \and
          Mario Radovich
          \inst{14}          
          \and
          Robert Reischke 
          \inst{3,4}          
          \and 
          Maximilian Von Wietersheim-Kramsta
          \inst{15,16}
          }

   \institute{Lorentz Institute for Theoretical Physics, Leiden University, PO Box 9506, NL-2300 RA Leiden, the Netherlands
         \and
         Leiden Observatory, Leiden University, PO Box 9513, NL-2300 RA Leiden, The Netherlands
         \and
         Universit\"at Bonn, Argelander-Institut f\"ur Astronomie, Auf dem H\"ugel 71, 53121 Bonn, Germany
         \and
         Ruhr University Bochum, Faculty of Physics and Astronomy, Astronomical Institute (AIRUB), German Centre for Cosmological Lensing, 44780 Bochum, Germany
         \and
         School of Mathematics, Statistics and Physics, Newcastle University, Herschel Building, NE1 7RU, Newcastle-upon-Tyne, UK
         \and
         Center for Theoretical Physics, Polish Academy of Sciences, al. Lotników 32/46, 02-668 Warsaw, Poland
         \and
         Institute for Astronomy, University of Edinburgh, Royal Observatory, Blackford Hill, Edinburgh, EH9 3HJ, UK
         \and
         Department of Physics and Astronomy, University College London, Gower Street, London WC1E 6BT, UK
         \and
         Zentrum für Astronomie, Universitatät Heidelberg, Philosophenweg 12, D-69120 Heidelberg, Germany
         \and
         Institute for Theoretical Physics, Philosophenweg 16, D-69120 Heidelberg, Germany
         \and
         Dipartimento di Fisica e Astronomia "Augusto Righi" - Alma Mater Studiorum Università di Bologna, via Piero Gobetti 93/2, I-40129 Bologna, Italy
         \and
         Istituto Nazionale di Astrofisica (INAF) - Osservatorio di Astrofisica e Scienza dello Spazio (OAS), via Piero Gobetti 93/3, I-40129 Bologna, Italy
         \and
         Istituto Nazionale di Fisica Nucleare (INFN) - Sezione di Bologna, viale Berti Pichat 6/2, I-40127 Bologna, Italy
         \and
         INAF - Osservatorio Astronomico di Padova, via dell'Osservatorio 5, 35122 Padova, Italy
         \and
         Institute for Computational Cosmology, Ogden Centre for Fundament Physics - West, Department of Physics, Durham University, South Road, Durham DH1 3LE, UK
         \and
         Centre for Extragalactic Astronomy, Ogden Centre for Fundament Physics - West, Department of Physics, Durham University, South Road, Durham DH1 3LE, UK
         }

   \date{Received XXX; accepted YYY \vspace{-0.1cm}}

   \abstract{Direct measurements of the matter power spectrum, $P_\mathrm{m}(k,z)$, provide a powerful tool to investigate observed tensions between models of structure growth while also testing the internal consistency of cosmological probes. We analyse cosmic shear data from the final data release of the Kilo-Degree Survey (KiDS), presenting a deprojected $P_\mathrm{m}(k,z)$, measured in up to three redshift bins. Compared to analyses using previous KiDS releases, we find improved internal consistency in the $z\lesssim0.7$ regime. At large scales, $k\lesssim0.1\,h\,\rm Mpc^{-1}$, our power spectrum reconstruction aligns with $\Lambda$CDM predictions with a density fluctuation amplitude $\sigma_8=0.81$. Furthermore, at small scales, $k=3$--$20\,h\,\rm Mpc^{-1}$, the average matter power spectrum is suppressed by $30\%\pm10\%\,{\rm (stat.)}\pm4\%\,{\rm (sys.)}$ with $2.8\sigma$ significance relative to a dark-matter-only model, consistent with expectations of strong baryonic feedback.}

   \keywords{Gravitational lensing: weak -- large-scale structure of Universe -- Cosmology: observations\vspace{-0.2cm}}

   \maketitle

\section{Introduction}
   
The growth of cosmic structure is quantified by the matter power spectrum, $P_\mathrm{m}(k,z)$, which is probed across different wavenumbers, $k$, and redshifts, $z$ by cosmological observables like cosmic microwave background lensing \citep[CMB; e.g.,][]{Lewis2006_CMBlensing}, galaxy clustering \citep[e.g.,][]{Reid2010_ggclustering}, the Lyman-alpha forest \citep[e.g.,][]{Rauch1998_Lya}, and weak gravitational lensing \citep[WL; e.g.,][]{Kilbinger2015_cosmicshear}. 

To identify the origins of discrepancies in cosmology inference between different surveys and probes operating at different scales or times, the shape and amplitude of $P_\mathrm{m}(k,z)$ can be inferred instead of condensing observations by physical model parameters \citep[][]{Amon2022_amod}. This approach identifies possible suppression signals relative to a power spectrum in a minimalistic $\Lambda$CDM scenario that accounts for structure growth by dark matter only (DMO). The isolated suppression signal may then be used to differentiate signatures from sources within known model uncertainties, such as baryonic effects, or new physics \citep[e.g.,][]{Preston2024_forcastpowerspec,Preston2025_DM_powspec}. In this work, we focus on the interpretation of weak lensing data, for which the uncertainty on the shape of $P_\mathrm{m}(k,z)$ is substantial. As such, a functional form is typically assumed for its shape, or redshift evolution, or both. Proposed models include either perturbative expansions \citep[][]{Ye2024_powspecreconstruct} or phenomenological models \citep[][]{1998ApJ...506...64S,2003MNRAS.346..994P,PerezSarmiento2025_powspecreconstr}.

In this paper, we combine the approaches of \cite{Broxterman2024_firststep}, hereafter BK24, \cite{Simon2012_Pmk_reconstruct}, and \cite{Simon2025_regularisation}, hereafter SPBK25, by adopting two different functional forms to extract direct measurements of the matter power spectrum from cosmic shear data. We compare a stiff, double power-law model for $P_{\rm m}(k,z)$ (BK24), with a flexible deprojected regularised model that infers deviations from a best-fitting DMO reference power spectrum (SPBK25). Both methods, applied to the fourth Kilo-Degree Survey data release \citep[][hereafter KiDS-1000]{Kuijken2019_kids1000}, show evidence of $P_\mathrm{m}(k,z)$ inconsistent with the $\Lambda$CDM constraints derived from the same data, specifically in the redshift evolution. SPBK25 discuss errors in the adopted redshift distributions within tomographic bins or the modelling of intrinsic alignment (IA) of sources as possible reasons for this inconsistency. Furthermore, the recent work from \citet{Doux2025} reports differences between Stage III lensing surveys and \textit{Planck} for $P_{\rm m}(k,z)$ that are in line with those found in BK24 and SPBK25.

The recently published final KiDS data release \citep[KiDS-Legacy;][]{Wright2024_KiDS_DR5}, featuring more area, an improved redshift calibration, one additional higher-redshift tomographic bin, and a new shear catalogue, supersedes KiDS-1000. While the previous KiDS-1000 results show a mild tension ($2\sigma$--$3.5\sigma$) with \textit{Planck} CMB measurements \citep[][]{Hildebrandt2017_KiDS,Asgari2021KiDS}, pointing towards a lower $S_8$-amplitude of $P_{\rm m}(k,z)$, this tension is now alleviated in the final cosmic shear analysis \citep{Stolzner2025_kids,Wright20251_cosmoresults}. This shift in findings motivates us to repeat our previous $P_\mathrm{m}(k,z)$ analysis using the new, higher-quality data.

This Letter presents the result of these KiDS-Legacy $P_\mathrm{m}(k,z)$ reconstructions with the BK24 and SPBK25 approaches. Section~\ref{sec:methods} summarises the data and describes minor methodological changes relative to   the previous works. Section \ref{sec:results} presents the deprojected matter power spectrum and ratios relative to the DMO reference for single or multiple redshift bins, and compares the results to predictions from state-of-the-art cosmological hydrodynamical simulations. The main results are summarised in Sect.~\ref{sec:conclusions}. Throughout, distances and wavenumbers are reported in comoving units for the Hubble constant $H_0=100\,h\,\rm km\,s^{-1}\,Mpc^{-1}$. \vspace{-0.2cm}

\section{Data and analysis}\label{sec:methods}

We analysed the KiDS-Legacy shear two-point correlation functions, $\xi_\pm^{(ij)}(\theta)$, between tomographic bins $i$ and $j$ within the angular range $\theta \in [2^\prime,300^\prime]$. This analysis incorporated six source redshift distributions, $n^{(i)}_{\rm s}(z)$ with $\int_0^\infty{\rm d}z\,n_{\rm s}^{(i)}(z)=1$, and the theoretical covariance matrix of $\xi_\pm^{(ij)}(\theta)$ uncertainties, based on \cite{Reischke2024_onecovariance}. The methodologies employed in this paper are detailed in BK24 and SPBK25. Here, we only discuss specific changes to the original setup and the data used. Appendix~\ref{app:model_details} provides additional details on the original methods. Compared to KiDS-1000, major improvements include updated angular scale cuts, new tomographic bins including an additional sixth tomographic bin extending to a photo-$z$ regime of $1.14 < z_{\rm B}\le2$ with deeper $i$-band imaging, a larger survey area ($1347\,\rm deg^2$, a $\sim 34\%$ extension), improved redshift distribution calibration methods \citep[][]{Wright20252_Legacycalibration} leveraging a larger calibration sample (five times the number of spectroscopic redshifts), new multiband image simulations for shear calibration \citep[][]{SSLi2023_Skills}, and a new IA model featuring redshift evolution. The combined increase in survey area and the extra tomographic bin leads to a roughly 3.5-fold increase in the probed cosmic volume.

The novel IA model is motivated by the observed dichotomy between red and blue galaxies, and the observed scaling of the IA strength with halo mass (e.g., \citealt{Fortuna2025_IA}, \citealt{Navarro2025}). Consistent with the KiDS-Legacy cosmological analysis, we therefore updated the IA kernel to
\begin{align}\label{eqn:IA_kernal}
  W^{(i)}_{\mathrm{I}}(\chi) =
  \underbrace{-A_{\mathrm{IA}}\, f_\mathrm{r}^{(i)}\,
  \bigg(\frac{\langle M_\mathrm{h} \rangle^{(i)}}{M_{\mathrm{h,pivot}}}\bigg)^\beta\,
  \frac{C_1\,\rho_{\mathrm{cr}}\,\Omega_{\mathrm{m}}}{D_+(\chi)}}_{=:F^{(i)}(\chi)}
  \,n^{(i)}_s(z[\chi])\,
  \frac{{\rm d}z(\chi)}{{\rm d}\chi}\;,
\end{align}
where $\chi$ is the comoving distance, $z(\chi)$ is the redshift at $\chi$, $A_\mathrm{IA}$ is the IA amplitude, $\Omega_\mathrm{m}$ is the total matter density relative to the critical density, $D_+(\chi)$ is the linear growth factor (defined such that $D_+(0)=1$), $C_1\,\rho_{\mathrm{cr}}=0.0134$ is a constant, $M_{\mathrm{h,pivot}} = 10^{13.5}\,h^{-1}\,M_\odot$ is the pivot halo mass, $\langle M_\mathrm{h} \rangle^{(i)}$ is the average halo mass in $z$-bin $i$, and $\beta$ is the power-law exponent denoting the evolution with halo mass. Our fractions of red galaxies, $f_\mathrm{r}^{(i)}$, use the values in table B.1 of \citet{Wright20251_cosmoresults}.

In our IA upgrade, the deprojected regularisation technique now employs $F^{(i)}(\chi)$ for bin $i$, instead of the previous $F(\chi)$, rendering the IA amplitude  $z$-dependent. Furthermore, we marginalised the posterior of $P_{\rm m}(k,z)$ over the distribution of $A_\mathrm{IA}$, $\beta$, and $\langle M_\mathrm{h} \rangle^{(i)}$ from the KiDS-Legacy cosmological analysis \citep{Wright20251_cosmoresults}. This was done after asserting the Gaussian prior in SPBK25 of \mbox{$\Omega_{\rm m}\sim{\cal N}(0.305,0.012)$} for the angular diameter distance in a flat $\Lambda$CDM universe. The DMO reference power spectrum, $P^{\rm DMO}_{\rm m}(k,z)$, against which deviations \mbox{$f_\delta(k,z):=P_{\rm m}(k,z)/P^{\rm DMO}_{\rm m}(k,z)$} are measured, was again \texttt{halofit} \citep{Takahashi2012_halofit} with the SPBK25 parameters, except for a higher \mbox{$\sigma_8=0.81$} to match the increased $\xi_\pm^{(ij)}(\theta)$ amplitude in KiDS-Legacy. We also marginalised over uncertainties in the source redshift distributions, for which we used the (shear-independent) error model from \cite{Wright20252_Legacycalibration} for the calibrated $n_{\rm s}^{(i)}(\chi)$.

For the double power-law approach, we fixed the IA parameters to the best-fitting $\Lambda$CDM values. This choice of IA model details does not have a significant impact: even in the extreme case where we impose $A_{\mathrm{IA}}=0$, our results typically vary by less than $5\%$--$10\%$, mostly at high $z$. 
\vspace{-0.2cm}

\section{Results and discussion}\label{sec:results}

We applied the regularised deprojection method (SPBK25) to constrain the deviations $f_\delta(k,z)$ in 20 logarithmic bins within \mbox{$k=0.01$--$20\, h\,\rm Mpc^{-1}$}. We used two settings: either averaging over the full redshift range $Z=[0,2]$, or over three separate redshift bins with boundaries $Z_1=[0,0.3]$, $Z_2=[0.3,0.6]$, and $Z_3=[0.6,2]$ to probe for redshift evolution. 
In the BK24 method, we fitted the double power-law to the full angular scale and redshift range of the $\xi_\pm^{(ij)}(\theta)$ data provided.
\vspace{-0.1cm}
\subsection{Matter power spectrum constraints}

\begin{figure}
  \centering
  \includegraphics[width=0.9\hsize]{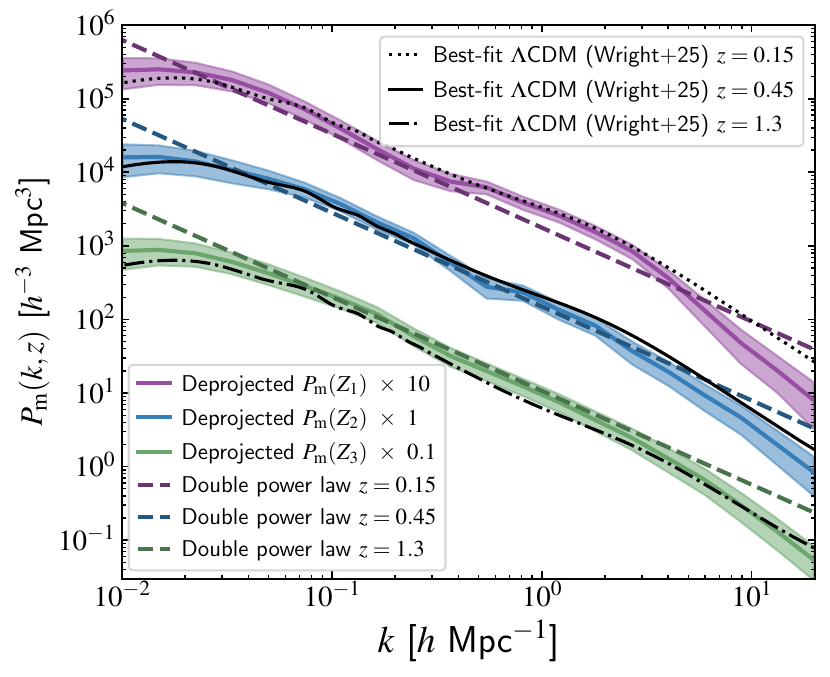}
  \caption{Power spectrum constraints in three variants. The solid coloured lines are regularised deprojections with $68\%$ CIs for three redshift bins $Z_1=[0,0.3]$, $Z_2=[0.3,0.6]$, and $Z_3=[0.6,2]$, and the dashed lines are best-fits of a double power-law, Eq. (\ref{eq:dpl_constraints}), all interpolated to the centres of $Z_1$--$Z_3$. The best-fitting $\Lambda$CDM constraints by \cite{Wright20251_cosmoresults} are shown as black curves. For clarity, the curves corresponding to different redshift bins are scaled by factors of 0.1, 1, or 10. Lensing constraints for $Z_3$ are mostly from structure near $z\sim0.7$ (see text). \vspace{-0.1cm}}
  \label{fig:matter_power_spectra}
\end{figure}

Figure~\ref{fig:matter_power_spectra} presents the inferred $P_\mathrm{m}(k,z)$ as a function of scale for the three redshift bins $Z_1$--$Z_3$. The Bayesian posterior constraints, $P_{\rm m}(k,z)=f_\delta(k,z)\,P^{\rm DMO}_{\rm m}(k,z)$, are plotted as solid curves for the median with 68th percentile credible intervals (CIs) at the central values, $z_{\rm c}$, of their respective redshift bins: $z_{\rm c}=0.15$ (purple), 0.45 (blue), and 1.3 (green); the posterior predictive distribution in comparison to the KiDS-Legacy data, and their good match, is shown in Appendix~\ref{app:ppd}. As pointed out in SPBK25, most of the lensing signal on the average $f_\delta(k,z)$ originates from lower redshifts within $Z_1$--$Z_3$, approximately $z=0.13$, $0.4$, and $0.7$, respectively. The dashed curves correspond to the best-fit double power-law (with $k_\mathrm{piv}=0.5~h~$Mpc$^{-1}$ and $z_\mathrm{piv} = 0.33$),
\begin{equation}\label{eq:dpl_constraints}
  P_{\rm m}(k,z)=
  10^{2.56^{+0.02}_{-0.02}}\,
  \left(\frac{1+z}{1+z_{\rm piv}}\right)^{-0.7^{+0.4}_{-0.4}}\,
  \left(\frac{k}{k_\mathrm{piv}}\right)^{-1.28^{+0.02}_{-0.02}}\!\!\!\!\!\!h^{-3}\,\mathrm{Mpc}^3\;,
\end{equation}
and the additional black curves in the figure correspond to the $\Lambda$CDM matter power spectrum obtained with the best-fitting parameters from the KiDS-Legacy cosmic shear inference.

In general, the two reconstruction methods and the $\Lambda$CDM fit provide consistent constraints for $P_{\rm m}(k,z)$, albeit within their limitations. For instance, while the regularised deprojection offers maximum flexibility (\mbox{$0\le f_\delta\le100$} in 20 $k$-bins), a strict power-law across the entire range $k=0.01$--$20\,h\,\rm Mpc^{-1}$ is not a good fit to $\Lambda$CDM lines at the edges of the $k$-range, specifically for $k\lesssim0.03\,h\,\rm Mpc^{-1}$ or $k\gtrsim5\,h\,\rm Mpc^{-1}$. But even the $\Lambda$CDM constraint exhibits excessive power in the non-linear regime (\mbox{$k\gtrsim3\,h\,\rm Mpc^{-1}$}) when compared to the posterior CIs of the regularised reconstruction. Regarding redshift evolution, the double power-law indicates a slightly weaker evolution with redshift than the regularised construction. However, the double power-law is also ill-constrained, as shown in Eq.~(\ref{eq:dpl_constraints}) and, due to deviations from a simple power-law, it also depends on the $k$-range used in the fit. Nevertheless, the detected amplitude change in the power-law improves upon the previous inconclusive KiDS-1000 results (BK24), clearly favouring structure growth with time. A similar conclusion can be drawn from the regularised deprojection, being consistent with the $\Lambda$CDM model and its change with $z$ for \mbox{$k\lesssim0.1\,h\,\rm Mpc^{-1}$}.
\vspace{-0.1cm}
\subsection{Internal consistency with $\Lambda$CDM}

\begin{figure}
  \centering
  \includegraphics[width=0.9\hsize]{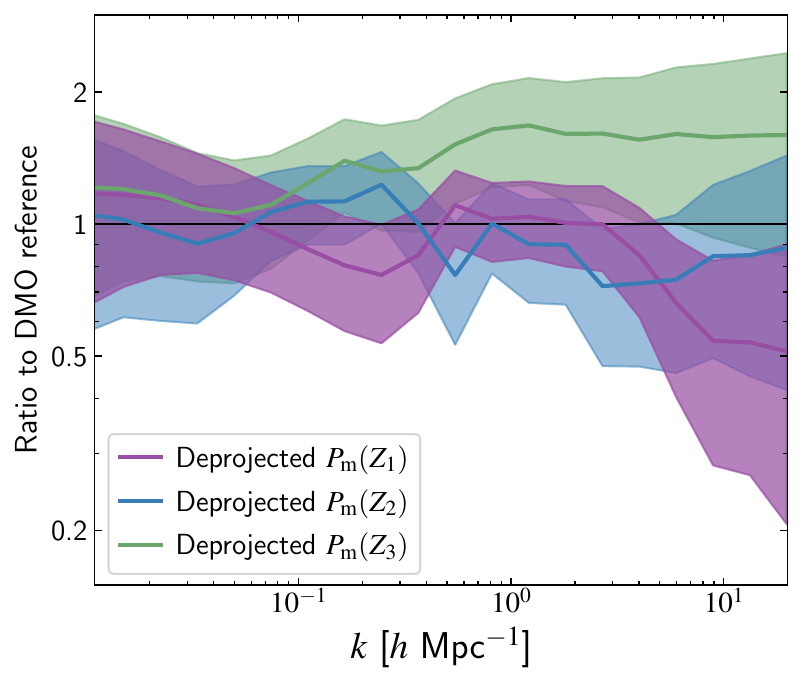}
  \caption{Average ratio of the matter power spectrum to the DMO reference within the three redshift bins $Z_1$--$Z_3$, derived using the method of regularised deprojection. The shaded regions represent the $68\%$ CIs of the posterior constraints for $Z_1$ (purple), $Z_2$ (blue), and $Z_3$ (green). At large scales, $k\lesssim0.1\,h\,\rm Mpc^{-1}$, all three redshift bins align with the expected DMO redshift evolution.\vspace{-0.1cm}}
  \label{fig:internal_consistency}
\end{figure}

In Fig.~\ref{fig:internal_consistency}, we present the constraints for $f_\delta(k,z)$ for three redshift bins ($Z_1$: purple, $Z_2$: blue, $Z_3$: green). The bins are individually $68\%$ CI consistent with their DMO reference for $k\lesssim0.1\,h\,\rm Mpc^{-1}$. At the $1\sigma$ level, however, the highest redshift bin, $Z_3$, still prefers slightly more power than the reference spectra at intermediate scales, $k\sim1\,h\,\rm Mpc^{-1}$, which might reflect an actual deviation from a DMO scenario on nonlinear scales. Our findings are in contrast with the analysis of SPBK25, which identified an internally inconsistent evolution of $P_\mathrm{m}(k,z)$ with the best-fitting $\Lambda$CDM cosmology. SPBK25 show that for KiDS-1000, the deviations at medium redshift, $Z_2$, and high redshift, $Z_3$, differ significantly from the KiDS-1000 DMO reference, resulting in a $P_{\rm m}(k,z)$ without significant structure growth between $z=0.4$--$0.7$, including at large scales. This inconsistency is no longer present in our analysis with KiDS-Legacy.

We attribute the improvement over KiDS-1000 primarily to the enhanced redshift-estimation methodology and the increase in data samples \citep{Wright20252_Legacycalibration}. This aligns with the conclusions of \citet{Wright20251_cosmoresults} and \citet{Stolzner2025_kids}, who identify the improved redshift distribution estimation as a major contributor to alleviating the $S_8$ tension. Further factors contributing to consistency with the \textit{Planck} cosmology include the increased survey volume and improved shear calibration. Our analysis also indicates that the refined IA model has a minor beneficial impact on the robustness for the deprojected $P_{\rm m}(k,z)$: it shifts the CI in Fig.~\ref{fig:internal_consistency} closer to $f_\delta(k,z)=1$ by $5\%$--$10\%$ relative to results without IA treatment (not shown). The potential for systematic errors in $f_\delta(k,z)$ stemming from redshift calibration errors and $z$-dependent IA are further discussed in SPBK25, but contribute less than $10\%$ to the total statistical error.
\vspace{-0.1cm}
\subsection{Small-scale suppression}

\begin{figure}
  \centering
  \includegraphics[width=0.9\hsize]{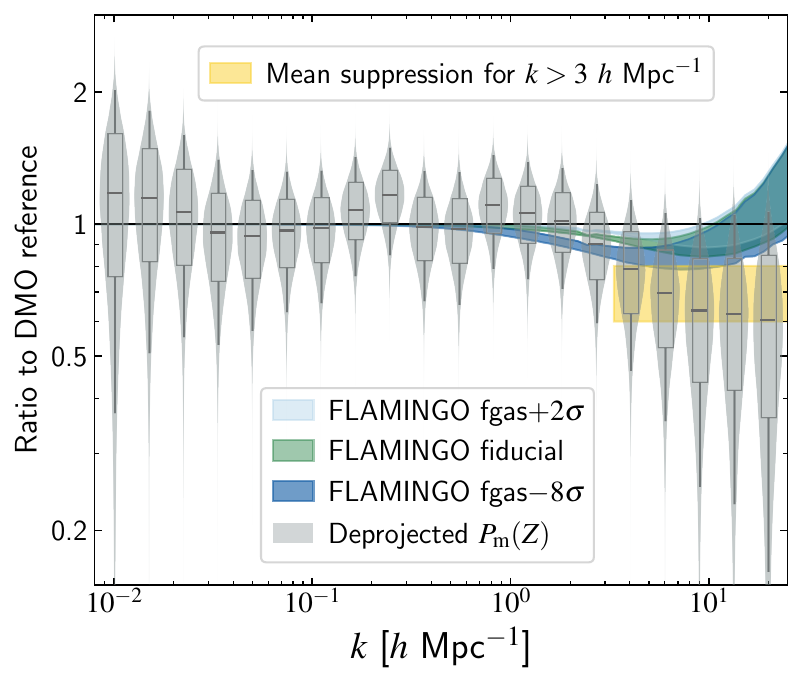}
  \caption{Violin plot of the matter power spectrum relative to the DMO spectrum in 20 bands, similar to Fig. \ref{fig:internal_consistency}, but for a single broad redshift bin, $Z$, which combines the smaller bins $Z_1$--$Z_3$. The width of the shaded regions represents the posterior probability density, with the 68th and 95th percentile CIs about the median also shown inside the regions as open boxes and sticks, respectively. The green, light blue, and dark blue bands illustrate the suppression predicted by the FLAMINGO cosmological hydrodynamical simulation between $z=0$ and 1.5 for the fiducial, weak, and strong feedback models, respectively. The yellow band indicates the mean suppression at small angular scales ($68\%$ CI), including correlations between the different points (Fig. \ref{fig:corr_mat}), preferring the strong feedback model. \vspace{-0.1cm}}
  \label{fig:single_bin}
\end{figure}

To boost the signal-to-noise ratio, the 20 grey violin data points in Fig.~\ref{fig:single_bin} average the constraints for $f_\delta(k,z)$ for a single broad redshift bin, $Z=Z_1\cup Z_2\cup Z_3$. The width of the contours represents the posterior probability density. The dark grey horizontal lines show the median, with the $68\%$ and $95\%$ CIs  indicated by open boxes and vertical black sticks. At the $1\sigma$ level, $f_\delta(k,z)$ is constrained to approximately $20\%$ for each $k$-bin. The broader CIs indicate that uncertainties increase towards the highest and smallest wavenumbers, where cosmic shear is less constraining. At small scales, $k\gtrsim3~h\,\mathrm{Mpc}^{-1}$, there is a clear suppression of the signal. This is better constrained by averaging all data points between $k=3$--$20\,h\,\mathrm{Mpc}^{-1}$. This average, which accounts for the error correlations of $r\lesssim0.3$ that increase for $k\gtrsim10\,h\,\rm Mpc^{-1}$ (Appendix~\ref{app:corr_mat}), yields $\bar{f}_\delta=0.70\pm0.10$. This indicates a $3\sigma$ significant detection of suppression of $P_\mathrm{m}(k,z)$ relative to $P^{\rm DMO}_{\rm m}(k,z)$. We verify this by randomly drawing $\bar{f}_\delta$ values from a null model, $f_\delta(k,z)\equiv1$, with errors as in our data; $0.3\%$ of these values have $\bar{f}_\delta\le0.70$, compatible with a $3\sigma$ detection. A similar fit to the results within the individual bins $Z_1$--$Z_3$ gives $\bar{f}_\delta=0.68\pm0.18$, $0.71\pm0.23$, and $1.51\pm0.57$ ($|r|\le0.4$), respectively. The increasing suppression with time is qualitatively consistent with cosmological hydrodynamical simulations \citep[e.g.,][]{Schaller2025_baryonsuprression} and tentatively suggests suppression only in $Z_1$ and $Z_2$ ($z\lesssim0.4$), with no suppression in $Z_3$ ($z\sim0.7$) at the $1\sigma$ level. In addition to statistical uncertainties, we anticipate a conservative systematic uncertainty of $\sim5\%$ for $\bar{f}_\delta$ because the \texttt{halofit} DMO reference deviates by a factor of $0.95$ to $1.05$ from the more accurate \texttt{hmcode2020} \citep{Mead2021_HMCode2020} in the regime of $k=2$--$10\,h\,\rm Mpc^{-1}$. Consequently, using \texttt{hmcode2020} as the DMO reference changes $\bar{f}_\delta$ by a similar factor.

For theoretical reference, the baryonic suppression signal from the FLAMINGO cosmological hydrodynamical simulations is over-plotted in Fig.~\ref{fig:single_bin} as shaded coloured areas \citep[]{schaye2023flamingo,kugel2023flamingo,Schaller2025_baryonsuprression}. These simulations are calibrated to (shifts in) the gas fractions of low-redshift groups and clusters. The green band in Fig.~\ref{fig:single_bin} corresponds to the fiducial FLAMINGO suppression between $z=0$--$1.5$. The light and dark blue bands represent variations calibrated to higher  (`$f_\mathrm{gas}+2\sigma$') and lower (`$f_\mathrm{gas}-8\sigma$') gas fractions, respectively, which covers the range of predicitions by nearly all modern cosmological simulations (see e.g., figure 11 of \citealt{Schaller2025_baryonsuprression}). The KiDS-Legacy suppression and the FLAMINGO predictions manifest in the same wavenumber regime, indicating that the signal is consistent with the expectation of baryonic feedback. This represents the most significant direct detection of baryonic feedback from cosmic shear alone (for previous measurements, see \citealt{Yoon2021_baryonsKids450} and \citealt{Chen2023_baryonsDESyr3}). Therefore, despite the deprojection and systematic uncertainty, \mbox{$\bar{f}_\delta=0.70\pm0.10\,{\rm (stat.})\pm0.04\,{\rm(sys.)}$}, there is a preference for the stronger feedback model in KiDS-Legacy. Individually, the weak, fiducial, and strong feedback FLAMINGO variations are, respectively, 2.6$\sigma$, 2.3$\sigma$, and 1.9$\sigma$ discrepant from our inferred value of $\bar{f}_\delta$. A stronger suppression resonates with recent studies that combine kinetic Sunyaev-Zeldovich (SZ) with either WL \citep{Bigwood2024_kSZ_WL}, clustering of photometric galaxies \citep{Hadzhiyska2024_kSZ_DESI}, galaxy-galaxy lensing \citep{McCarthy2025}, or X-ray \citep{2025arXiv250707991K}, as well as with analyses using thermal SZ and WL \citep{Troster2022,Pandey2025} or fast radio bursts \citep{reischke2025}.

Compared to our regularised deprojection, the KiDS-Legacy results by \citet{Wright20251_cosmoresults} prefer weaker feedback, at the level of $f_\mathrm{gas}+2\sigma$ or less (68\% CI), for \mbox{$k\lesssim3\,h\,\rm Mpc^{-1}$} (their figures 15 and F.3).  This discrepancy is also visible in Fig.~\ref{fig:matter_power_spectra}, where the KiDS-Legacy results (black lines using \texttt{hmcode2020}) favour greater power for \mbox{$k\gtrsim1\,h\,\rm Mpc^{-1}$} than our deprojected $P_{\rm m}(k,z)$, especially for $Z_1$ when extrapolated to $10\,h\,\rm Mpc^{-1}$. We attribute this difference primarily to the employed statistics and their sensitivity to different scales, given the choice of angular cuts. As shown in \citet[figure 1]{Wright20251_cosmoresults}, our $\xi^{(ij)}_\pm(\theta)$  are sensitive to \mbox{$\ell\sim10^4$}, whereas their COSEBIs with orders $n\le6$ strongly downweight signals beyond \mbox{$\ell\sim10^3$}. Therefore, compared to the full physics model and the a priori assumptions in \citet{Wright20251_cosmoresults}, our agnostic deprojection is sensitive to physics around \mbox{$k\sim5\,h\,\rm Mpc^{-1}$}, although it is also more susceptible to systematic errors within this nonlinear regime. This interpretation of a sensitivity shift aligns with section 4.3 of \citet{Stolzner2025_kids}, who find that band powers prefer stronger feedback, peaking near the upper edge of the KiDS-Legacy feedback parameter $T_{\rm AGN}$ prior. The combination of COSEBIs and bandpowers -- which have a window function similar to $\xi^{(ij)}_\pm(\theta)$ -- peaks at \mbox{$\log_{10}{(T_{\rm AGN}/\rm K)} \approx 8.0$}. This is, however, dominated by a Bayesian prior that explicitly excludes our suppression level.
\vspace{-0.2cm}

\section{Conclusions}\label{sec:conclusions}
Compared to KiDS-1000, the cosmic shear data from KiDS-Legacy show an improved degree of internal consistency in the context of $\Lambda$CDM. In particular, the deprojected growth of structure in the matter power spectrum is $1\sigma$ consistent with that of the $\Lambda$CDM inference on large scales (Fig.~\ref{fig:internal_consistency}). The change is primarily attributed to an improved source redshift calibration and the revised image reduction.

The data also reveal deviations from a DMO model on small scales, now probed to higher significance due to the $\sim3.5$ times larger survey volume. We measure growth suppression in the matter power spectrum relative to a DMO model of $1-\bar{f}_\delta=30\%\pm10\%\,{\rm (stat.)}\pm4\%\,{\rm (sys.)}$ at $k\ge3\,h\,\rm Mpc^{-1}$, with tentative evidence that this suppression is restricted to $z\lesssim0.4$. The average suppression, detected at $2.8\sigma$ significance from cosmic shear data alone, is consistent with baryonic feedback predictions in the FLAMINGO simulations and shows a preference for the stronger feedback variation (Fig.~\ref{fig:single_bin}).

\section{Data availability}

Our constraints for the deprojected $P_{\rm m}(k,z)$ and the correlation matrix of its uncertainties are made available at CDS. \vspace{-0.1cm}

   \bibliographystyle{aa} % style aa.bst
   \bibliography{references.bib} % your references Yourfile.bib

\begin{appendix}

\section{Acknowledgements}
LP acknowledges support from the DLR grant 50QE2302. HHi is supported by a DFG Heisenberg grant (Hi 1495/5-1), the DFG Collaborative Research Center SFB1491, an European Research Council (ERC) Consolidator Grant (No. 770935), and the DLR project 50QE2305. MA acknowledges support from the UK Science and Technology Facilities Council (STFC) under grant number ST/Y002652/1 and the Royal Society under grant numbers RGSR2222268 and ICAR1231094. MB is supported by the Polish National Science Center through grants no. 2020/38/E/ST9/00395 and 2020/39/B/ST9/03494. CH acknowledges support from the Max Planck Society and the Alexander von Humboldt Foundation in the framework of the Max Planck-Humboldt Research Award endowed by the Federal Ministry of Education and Research, and the UK STFC under grant ST/V000594/1. HHo acknowledges support from the ERC under the European Union’s Horizon 2020 research and innovation program with Grant agreement No. 101053992. BJ acknowledges support by the ERC-selected UKRI Frontier Research Grant EP/Y03015X/1 and by STFC Consolidated Grant ST/V000780/1. SSL is receiving funding from the programme ``Netzwerke 2021'', an initiative of the Ministry of Culture and Science of the State of Northrhine Westphalia. SSL acknowledges support from the European Research Council (ERC) under the European Union’s Horizon 2020 research and innovation program with Grant agreement No. 101053992. LM acknowledges the financial contribution from the grant PRIN-MUR 2022 20227RNLY3 “The concordance cosmological model: stress-tests with galaxy clusters” supported by Next Generation EU and from the grant ASI n. 2024-10-HH.0 “Attività scientifiche per la missione Euclid – fase E”. RR is supported by an ERC Consolidator Grant (No. 770935). MvWK acknowledges the support by UK STFC (grant no. ST/X001075/1), the UK Space Agency (grant no. ST/X001997/1), and InnovateUK (grant no. TS/Y014693/1). Based on data obtained from the ESO Science Archive Facility with DOI: \url{https://doi.org/10.18727/archive/37}, and \url{https://doi.eso.org/10.18727/archive/59} and on data products produced by the KiDS consortium. The KiDS production team acknowledges support from: Deutsche Forschungsgemeinschaft, ERC, NOVA and NWO-M grants; Target; the University of Padova, and the University Federico II (Naples). This work used the DiRAC@Durham facility managed by the Institute for Computational Cosmology on behalf of the STFC DiRAC HPC Facility (\url{www.dirac.ac.uk}). The equipment was funded by BEIS capital funding via STFC capital grants ST/K00042X/1, ST/P002293/1, ST/R002371/1, and ST/S002502/1, Durham University and STFC operations grant ST/R000832/1. DiRAC is part of the National e-Infrastructure.

\section{Method details}\label{app:model_details}

In this Appendix, we provide additional information on the method of regularised deprojection in SPBK25 and that of the double power-law fit in BK24.

\subsection{Regularised deprojection}
\renewcommand{\d}[0]{{\rm d}}

For a full account of the regularised reconstruction, we refer the reader to SPBK25, sections 2 and 3. The method's aim is to invert, without a physical model for the matter power spectrum, the relations
\begin{multline}
  \label{eq:xipm_pure}
  \xi_\pm^{(ij)}(\theta)=\frac{9H_0^4\Omega_{\rm m}^2}{4c^4}\\
  \times\int_0^{\chi_{\rm h}}
  \!\!\!\!\int_0^\infty\frac{\d\chi\,\d\ell\,\ell}{2\pi}
  \frac{\overline{W}^{(i)}\!(\chi)\overline{W}^{(j)}\!(\chi)}{a^2(\chi)}
  J_{0,4}(\ell\theta)\, P_{\rm
    m}\left(\frac{\ell+1/2}{f_{K}(\chi)},z[\chi]\right)
\end{multline}
with respect to $P_{\rm m}(k,z)$ for a set of tomographic bins
$i,j=1\ldots N_s$. Here, the lensing efficiency is defined as
\begin{equation}
  \label{eq:lenseff}
  \overline{W}^{(i)}\!(\chi):=
  \int_\chi^{\chi_{\rm h}}\d\chi^\prime\;
  n_{\rm s}^{(i)}(z[\chi])\,\frac{\d z(\chi)}{\d\chi}\,
  \frac{f_{K}(\chi^\prime-\chi)}{f_{K}(\chi^\prime)}\;, 
\end{equation}
which varies with each source distribution $n_{\rm s}^{(i)}(z)$, where $c$ is the vacuum speed of light, $H_0$ is the Hubble parameter, and $\chi_{\rm h}$ is the horizon of the survey. For brevity, we ignore here two additional IA terms, which, under the adopted IA model, only change the projection kernels inside the $N_s$ integrals (\ref{eq:xipm_pure}) and do not impact the subsequent statements. The projection kernels depend exclusively on $n_{\rm s}^{(i)}(z)$, the comoving angular diameter distance
$f_K(\chi)$ (for the curvature scalar $K$ and the comoving distance $\chi$), the matter density $\Omega_{\rm m}$, and the IA parameters. Uncertainties in these four components are propagated into $P_{\rm m}(k,z)$ by varying the projection kernels in repeated reconstructions.

An exact inversion is infeasible due to a low signal-to-noise ratio, confined angular ranges, and the limited number
$N_s\,(N_s+1)/2$ of equations in (\ref{eq:xipm_pure}). Therefore, as approximation, we average $P_{\rm m}(k,z)$ within $N_z=1$ or 3 broad $z$-bins in radial direction, spanning $z=0$--$2$, and within $N_k=20$ $k$-bands, ranging between $0.01$--$20\,h\,\rm Mpc^{-1}$. This configuration probes either the full radial average, $N_z=1$, or the evolution within $N_z=3$ bins. Outside these $k$ and $z$ ranges, the power spectrum is fixed to a reference power spectrum, $P^{\rm DMO}_{\rm m}(k,z)$, best-fit to the data ($\sigma_8=0.81$). All contributions from outside these ranges are encapsulated in the modelled shear signal $\xi^{(ij)}_{\pm,\rm fid}(\theta)$, which constitutes only a small fraction of the expected total signal $\xi^{(ij)}_\pm(\theta)$.  Within these ranges, the amplitudes in the bands can vary freely between a factor 0 to 100 of $P^{\rm DMO}_{\rm m}(k,z)$.

However, broad $z$-bins complicate the interpretation of the averaged power spectrum due to its variation with $z$. We untangle this by expressing $P_{\rm m}(k,z)=f_\delta(k,z)\,P^{\rm DMO}_{\rm m}(k,z)$ relative to the growing $P^{\rm DMO}_{\rm m}(k,z)$, and then average the more slowly evolving deviations, $f_\delta(k,z)$, instead. Since the adopted DMO reference accounts for dark matter only, $f_\delta(k,z)$ also conveniently averages the deviations of the true $P_{\rm m}(k,z)$ from a model without baryon feedback. Mathematically, our statistical model expresses the deviations $f_\delta(k,z)$ across $N_z\times N_k$ bands as the vector $\vec{f}_\delta$. With the linear projections (\ref{eq:xipm_pure}) presented by the projection matrix $\tens{X}$, the predicted $\theta$-binned $\xi^{(ij)}_\pm(\theta)$ is given by $\vec{\xi}=\tens{X}\vec{f}_\delta+\vec{\xi}^{\rm fid}$ (assuming fixed projection kernels).

The inversion of the noisy $\vec{\xi}=\tens{X}\vec{f}_\delta+\vec{\xi}^{\rm fid}$ without further constraints leads to strongly oscillating noise for $\vec{f}_\delta$ due to the broad projection kernels and the Bessel functions of first kind, $J_n(x)$, inside the kernels. This issue is mitigated by asserting positive solutions, $f_\delta(k,z)\ge0$, and by employing a Tikhonov filter (as in SPBK25).  This filter attenuates solutions of the band-power $\vec{f}_\delta$ that exhibit strong oscillations in the $k$-direction, effectively reconstructing a $k$-smoothed solution. Conversely, oscillations in $z$-direction for \mbox{$N_z>1$} are not filtered to prevent $z$-smoothing. SPBK25 uses simulated KiDS data for filter settings that strike a balance between degradation by noise and artificial smoothing by the filter.

We infer the band-power $\vec{f}_\delta$ statistically within a Bayesian framework using the posterior probability density function (PDF)
\mbox{$P(\vec{f}_\delta|\vec{\xi})\propto{\cal
    L}(\vec{\xi}|\vec{f}_\delta)\,P_{\rm
    hat}(\vec{f}_\delta)\,P_\tau(\vec{f}_\delta)$}. Here, the prior
PDFs $P_{\rm hat}(\vec{f}_\delta)$ and $P_\tau(\vec{f}_\delta)$ implement a uniform prior density, asserting $f_\delta=0$--$100$, and the Tikhonov regularisation of $\vec{f}_\delta$. The likelihood function, ${\cal L}(\vec{\xi}|\vec{f}_\delta)$, adopts a multivariate Gaussian model for statistical noise in $\vec{\xi}$ with covariance matrix $\tens{C}$. This covariance incorporates contributions from intrinsic shape noise, cosmic variance, super-sample covariance, and the uncertainty in the multiplicative shear bias. We numerically sample $P(\vec{f}_\delta|\vec{\xi})$ using a rapidly converging Hamiltonian Monte Carlo algorithm. To propagate statistical errors in $n_{\rm s}^{(i)}(z)$, the angular diameter distance and $\Omega_{\rm m}$, and IA correlations, we combine different Monte Carlo chains where the projection kernels are randomly drawn from a model of kernel uncertainties. The inferred
$P_{\rm m}(k,\bar{z})=\bar{f}_\delta(k)\,P^{\rm DMO}_{\rm  m}(k,\bar{z})$ at the centres, $\bar{z}$, of three redshift bins for the averaged $\bar{f}_\delta(k)$ is shown in Fig. \ref{fig:matter_power_spectra} with statistical uncertainties.

\subsection{Double power-law}
For the shape of the matter power spectrum, the analysis of BK24 assumes a double power-law of the form
\begin{equation}
    P_\mathrm{m}(k,a) = A\, \bigg(\frac{k}{k_\mathrm{piv}}\bigg)^p \bigg(\frac{a}{a_\mathrm{piv}}\bigg)^m,
\end{equation}
where the pivot points $k_\mathrm{piv}$ and $a_\mathrm{piv}$ are determined by minimising the covariance between the different model parameters. The free parameters $\log_{10}{(A/h^{-3}\, \mathrm{Mpc}^3)}$, $p$, and $m$ are inferred through a Bayesian inference by assuming the flat priors $\log_{10}{(A/h^{-3}\,\mathrm{Mpc}^3)} \in [0,10], p \in [-2,0.5],$ and $m \in [-5,5]$.

Additional freedom in modelling is chosen to replicate the choices made in the Stage III cosmic shear inferences. As such, the free parameters of the IA model, except for $A_\mathrm{IA}$, the integration limits of the different two-point statistics, and the scale cuts are taken from the fiducial Stage III cosmic shear inferences. The background cosmology, used to compute the linear growth factor and redshift-comoving distance relation, however, is chosen to be the best-fitting $\Lambda$CDM cosmology from \citet{PlanckVI2020}. Changing this to the best-fitting KiDS-1000 cosmology from \citet{Asgari2021KiDS} has a negligible impact on the results. 

As shown in Fig. \ref{fig:matter_power_spectra}, the maximum-likelihood solutions (dashed lines) are a reasonable fit to the regularised reconstruction within $0.03\lesssim k\lesssim 5\,h\,\rm Mpc^{-1}$. The systematically lower amplitude of the fit in this $k$-regime, compared to the regularised construction, is explained by increasing deviations from self-similar evolution outside this regime.

\section{Correlation matrix of deprojection uncertainties for a single redshift bin} \label{app:corr_mat}

\begin{figure}
  \centering
  \includegraphics[width=\hsize]{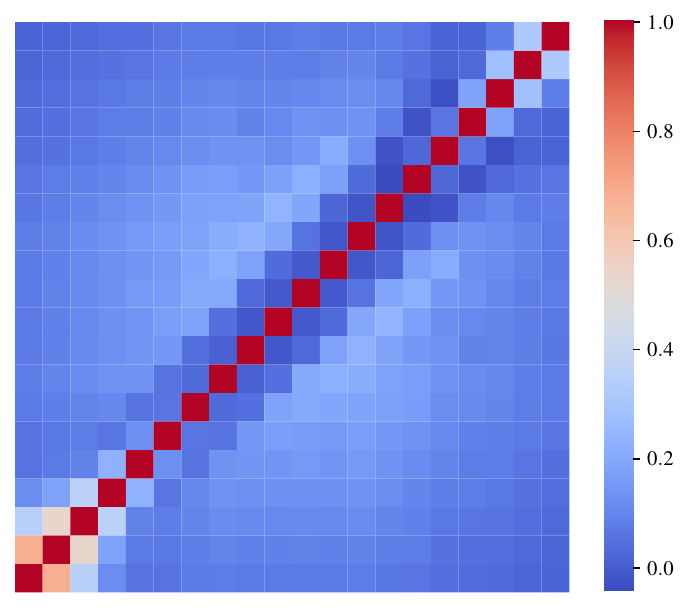}
  \caption{Matrix of Pearson correlation coefficients for the deprojection with a single $z$-bin, $Z$,
  for in 20 logarithmic $k$-bins between $k=0.01-20~h~\rm Mpc^{-1}$. The scale $k$ increases from the bottom left to the top right.}
  \label{fig:corr_mat}
\end{figure}

The deprojection of the tomographic $\xi_\pm^{(ij)}(\theta)$ into $f_\delta(k,z)$, with measurement noise localised in $\theta$, produces correlated uncertainties across all $k$ and $z$ bins in $f_\delta(k,z)$ and, consequently, in $P_{\rm m}(k,z)$ (SPBK25). Figure~\ref{fig:corr_mat} displays the correlation matrix for the results in Fig.~\ref{fig:single_bin} with single $z$-bin $Z$. The Pearson correlation coefficients, $r$,  of the deprojected $P_\mathrm{m}(k,z)$ between 20 logarithmically spaced bins for $k=0.01$--$20~h~\rm Mpc^{-1}$ are presented (from bottom left to top right). The blue colours indicate that estimates at different scales are at most moderately correlated, typically with $|r|\lesssim0.3$. The smallest and largest scales exhibit the strongest correlation with their directly adjacent bins. The correlation matrix for the deprojection with three separate $z$-bins ($Z_1$ to $Z_3$) is similar to that presented in figure 5 of SPBK25.

\section{Posterior predictive distribution}
\label{app:ppd}

\begin{figure*}
  \centering
  \includegraphics[width=\hsize]{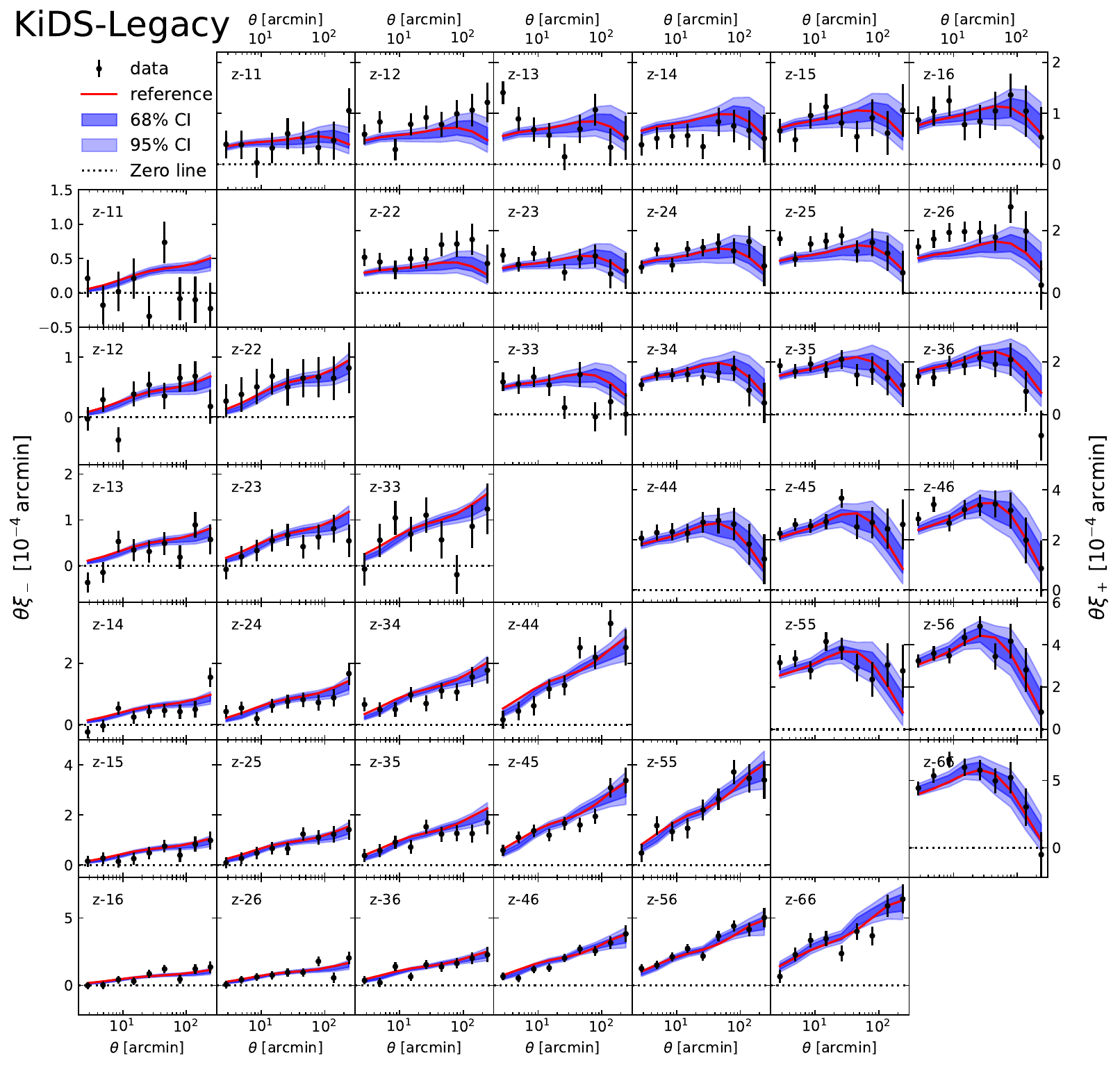}
  \caption{Posterior predictive distribution ($68\%$ and $95\%$ CIs as blue regions) of the deprojected $P_{\rm m}(k,z)$ in Fig. \ref{fig:matter_power_spectra} relative to the black data points -- $\theta\,\xi_-^{(ij)}(\theta)$ in lower left panels and $\theta\,\xi_+^{(ij)}(\theta)$ in the upper right panels -- for a combination $(ij)$ of tomographic source bins, denoted as ``$z-ij$'' inside the panels. The red lines represent the reference model, $P^{\rm DMO}_{\rm m}(k,z)$, employed in the regularised deprojection.}
  \label{fig:ppd}
\end{figure*}

Figure~\ref{fig:ppd} is the posterior predictive distribution (PPD) of $P_{\rm m}(k,z)$ deprojected into three separate redshift bins $Z_1$--$Z_3$ (solid coloured curves in Fig. \ref{fig:matter_power_spectra}). The $68\%$ and $95\%$ CIs of the PPD in blue assume a fixed lensing kernel and no IA uncertainties, which, if accounted for, modestly increase the CIs by approximately $5\%$. The solid red line represents the prediction from the DMO reference $P^{\rm DMO}_{\rm m}(k,z)$, adjusted in the Bayesian deprojection by a factor $f_\delta(k,z)$, presented in Fig. \ref{fig:internal_consistency}, to match the observed data (correlated black data points with $1\sigma$ uncertainty). Despite including an additional sixth source redshift bin compared to the previous KiDS-1000 analysis in SPBK25, our PPD consistently provides a good description of all $\xi^{(ij)}_\pm(\theta)$ data points, although a few notable outliers are present, such as for $\xi_\pm^{(33)}(\theta)$, $\xi_+^{(13)}(\theta)$, $\xi_-^{(12)}(\theta)$, or $\xi_+^{(26)}(\theta)$. 

\end{appendix}

\end{document}